\documentclass[a4paper, 10pt, conference]{ieeeconf}      

\IEEEoverridecommandlockouts                              
\usepackage[T1]{fontenc}
\usepackage[utf8]{inputenc}
\usepackage{graphicx}
\usepackage[fleqn]{amsmath}
\usepackage{amssymb}
\usepackage{multirow}
\usepackage{subfigure}
\usepackage{bm}
\usepackage{multicol}
\graphicspath{{./Figures/}}
\usepackage[hyperindex=true, %
pdftitle={Gaussian Process-Based Model Predictive Control of Blood Glucose for Patients with Type 1 Diabetes Mellitus},%
pdfauthor={Lukas Ortmann, Dawei Shi, Eyal Dassau, Francis J. Doyle III, Steffen Leonhardt, Berno J.E. Misgeld},%
colorlinks=true,%
pagebackref=false,%
plainpages=false,%
pdfpagelabels,%
linkcolor=black,%
citecolor=black,%
filecolor=black,%
urlcolor=black%
]{hyperref} 
\usepackage[dvipsnames]{xcolor}
\usepackage[absolute,overlay]{textpos}
\usepackage{url}
\title{\LARGE \bf Gaussian Process-Based Model Predictive Control of Blood Glucose\\ for Patients with Type 1 Diabetes Mellitus} 

\author{Lukas Ortmann$^{1}$, Dawei Shi$^{2}$, Eyal Dassau$^{2}$, Francis J. Doyle III$^{2}$, Steffen Leonhardt$^{1}$, Berno J.E. Misgeld$^{1}$
\thanks{This work was partially funded by the German Academic National Foundation and by the United States National Institutes of Health under grants DP3DK104057 and UC4DK108483.} 
\thanks{$^{1}$Philips Chair for Medical Information Technology, 
	Helmholtz-Institute, RWTH Aachen University, 52074 Aachen, Germany.
	{\tt\small lukas.ortmann@rwth-aachen.de}}
\thanks{$^{2}$Harvard John A. Paulson School of Engineering \& Applied Sciences, Harvard University, Cambridge, MA 02138, USA.}
}
\begin{document}
	\maketitle
	
	\begin{textblock*}{\textwidth}(15mm,10mm) 
		\centering \bf \textcolor{NavyBlue}{Published on \emph{Asian Control Conference (ASCC),} December 2017.\\\url{https://doi.org/10.1109/ASCC.2017.8287323}}
	\end{textblock*}
	\thispagestyle{empty}
	\pagestyle{empty}
	\renewcommand{\baselinestretch}{1}
	
	\begin{abstract} \label{Sec:Abstract}               
		The insulin sensitivity (IS) of the human body changes with a circadian rhythm. This adds to the time-varying feature of the glucose metabolism process and places challenges on the blood glucose (BG) control of patients  with Type 1 Diabetes Mellitus. This paper presents a Model Predictive Controller that takes the periodic IS into account, in order to enhance BG control. The future effect of the IS is predicted using a machine learning technique, namely, a customized Gaussian Process (GP), based on historical training data. The training data for the GP is continuously updated during closed-loop control, which enables the control scheme to learn and adapt to intra-individual and inter-individual changes of the circadian IS rhythm. The necessary state information is provided by an Unscented Kalman Filter. The closed-loop performance of the proposed control scheme  is evaluated for different scenarios (including fasting, announced meals and skipped meals) through \emph{in~silico} studies on simulation models of G\"ottingen Minipigs.
	\end{abstract}
	\smallbreak
	
	\section{Introduction}\label{Sec:Introduction}
	
	The number of people with diabetes mellitus (DM) worldwide was estimated to be around 415 million in 2015 and is predicted to increase to 642 million by the year 2040 \cite{Ogurtsova2017}. Approximately 10\% of them have Type 1 DM  (T1DM), which means they always need exogenous insulin to regulate their blood glucose (BG). Improper control of their BG concentration can lead to hyperglycemia ($\textrm{BG}>$ 180 mg/dL), induced secondary complications, and hypoglycemia ($\textrm{BG}<$ 70 mg/dL) induced by over delivery of insulin that can lead to life-threatening medical conditions. T1DM patients usually need to control their BG manually by measuring their BG concentration several times a day and take meal bolus insulin infusions to compensate for food intake in addition to continuous basal rate injections.
	Developments in continuous glucose monitoring sensors and insulin pumps, however, provide the opportunity of achieving improved blood glucose regulation via closed-loop control, which has been the goal of artificial pancreas device \cite{Thabit2016}, \cite{Doyle2014}.
	
	The glucose metabolism is subject to multiple influential factors that change in time (e.g. the quasi-periodic appearance of meals \cite{Wang2010}, the diurnal insulin sensitivity (IS) changes \cite{Toffanin2013}, and the irregular pattern of exercises \cite{Zinman2003}), which require the adoption of adaptive techniques in glucose regulation.
	A number of different controllers have been proposed in the literature to address these requirements.
	In \cite{Palerm2007}, the authors adapted the carbohydrate (CHO) to insulin ratio, which is used to calculate meal boluses, based on the patient's reaction to boluses given during the last days to learn and adapt to their personal metabolic system to improve the disturbance rejection against food intake.
	A similar approach of run-to-run control was used in \cite{Palerm2008} to update the basal infusion rate of patients based on past measurements to improve the tracking performance and to control the BG concentration to a tighter zone. In all of these studies, only a few BG measurements were available during the day.
	A Model Predictive Iterative Learning Controller was proposed in \cite{Wang2010} based on continuous glucose measurements to exploit the periodic appearance of meals to improve the control performance.
	The work in \cite{Toffanin2017} updated the basal rate during night time and the carbohydrate-to-insulin ratio (used to calculate meal boluses) during the day by using run-to-run adaptation. A clinical study on pigs, where both insulin and the counterregulatory hormone glucagon were used in a Generalized Predictive Control approach, was presented in \cite{El-Khatib2007}.
	An adaptive controller using Gain Scheduling based on the BG concentration was developed in \cite{Colmegna2016}, whereas in \cite{Misgeld2016b} an estimate of the IS determines which controller was activated.
	In \cite{Toffanin2013}, the authors included the circadian IS of their metabolic model into an input constraint of a Model Predictive Controller (MPC) to enhance the control performance.
	A review of adaptive controllers for BG regulation, including Minimum Variance Control, Self-Tuning-Regulators, Linear Quadratic Regulators and Generalized Predictive Control, can be found in \cite{Turksoy2014}.
	
	In this work, we develop a different adaptive control approach for glucose regulation. We focus on learning the effect of the individual's circadian IS and its changes during closed-loop control using continuous BG measurements to improve the tracking performance. We propose a combination of MPC, Unscented Kalman Filter (UKF) and Gaussian Process (GP) to learn the effect of the circadian IS rhythm on the metabolic system and incorporate the information in our controller. Through continuous learning during closed-loop control, the proposed controller method adapts to an individual. It enables personalized control and aims at solving the problems arising from large inter-individual and intra-individual changes in the IS of the glucose metabolism. This enables us to improve the tracking performance and control the BG concentration to a tighter zone, which corresponds to a smaller risk of secondary complications. The obtained \emph{in~silico} results on a mathematical model of G\"ottingen Minipigs indicate that the proposed controller is able to compensate for higher insulin resistance in the morning (known as ``dawn syndrome'') and brings the BG concentrations closer to the reference before breakfast time. By utilizing a detailed plant model and the UKF from \cite{Misgeld2017}, announced food intakes do not disturb the proposed learning algorithm. Skipped meals also do not affect the learning and do not disturb the controller performance. The time and amount of food intake can therefore be arbitrary and a fixed eating schedule is not necessary.
	
	The paper is organized as follows: In Section \ref{Sec:Modelling} we present the model used for the \emph{in~silico} studies and the controller design. Section \ref{Sec:Controller} introduces the proposed control algorithm including UKF, GP and MPC. In Section \ref{Sec:Results} we present and analyse the results of the \emph{in~silico} studies. Finally, Section~\ref{Sec:Conclusion} provides some concluding remarks.
	
	\section{System Modelling}\label{Sec:Modelling}
	
	A few models have been proposed in the literature to describe the glucose metabolism, e.g., the Bergman model \cite{Bergman1989}, Dalla Man model \cite{Dalla2014}, Hovorka model \cite{Hovorka2004} and Sorensen model \cite{Sorensen1985}. In \cite{Lunze2014b}, the Sorensen model was adapted to model the metabolic system of G\"ottingen Minipigs and was parametrized using data from animal trails. This model includes a linear parameter dedicated to the IS and thus enables us to easily incorporate a changing IS in the proposed control algorithm. Furthermore, the metabolic system of humans and pigs are quite similar \cite{Lunze2014a} and results obtained with this model can be migrated to a human glucose metabolic model. Due to space limitation, we only give a brief description of the G\"ottingen Minipig model. For implementation purpose, the controller is designed based on a linearization of this model.

	\subsection{Nonlinear G\"ottingen Minipig Model}
	
	The G\"ottingen Minipig model consists of 16 states to describe 9 different compartments. These compartments are used to model the glucose, insulin and glucagon concentration in the plasma (heart, brain, lung, kidneys, gut), the liver and the muscle/adipose tissue. They are interconnected through the blood stream, which links them to the gastro-intestinal tract and the subcutaneous insulin injections. The inputs to the system are subcutaneous insulin $u(t)$ and oral glucose $D_\textrm{{oral}}(t)$ and the output is the BG concentration $y(t)$.
	
	In the G\"ottingen Minipig model, the IS is incorporated through a linear parameter $k_{\textrm{IS}}$, which determines the effect of the insulin concentration in the interstitial space of muscular/adipose tissue $I_\textrm{{MI}}(t)$ on the glucose concentration in the intracellular fluid space of muscular/adipose tissue $G_\textrm{{MI}}(t)$ through
	\begin{equation}\label{eq:MiniPig_k_IS}
	\dot{G}_\textrm{{MI}}(t)=\left(G_\textrm{{MV}}(t)-G_\textrm{{MI}}(t)\right)/T_\textrm{{M}}^\textrm{{G}}-k_{\textrm{IS}}V_\textrm{{MI}}^\textrm{I}I_\textrm{{MI}}(t)/V_\textrm{{MI}}^\textrm{{G}},
	\end{equation}
	where $G_\textrm{{MV}}(t)$ is the glucose concentration in the vascular blood space of the muscular/adipose tissue, $T_\textrm{{M}}^\textrm{{G}}$ is a time constant and $V_\textrm{{MI}}^\textrm{{I}}$ and $V_\textrm{{MI}}^\textrm{{G}}$ are the distribution volumes of insulin and glucose in the interstitial space and intracellular fluid space of muscular/adipose tissue, respectively.
	In our engineering approach, we model the circadian IS of the glucose metabolism by making the corresponding linear parameter $k_{\textrm{IS}}$ time-dependent. The circadian rhythm of this parameter is built on and adapted from the periodic function obtained in \cite{Toffanin2013} (see Fig. \ref{fig:fasting}D). The modified IS has a nominal value of 1 mg/(min$\cdot$mU) with a low of 0.55 mg/(min$\cdot$mU) and a high of 1.4 mg/(min$\cdot$mU). This function will be used as an example rhythm in the \emph{in~silico} studies. Note that other periodic functions can also be learned by the proposed method, which enables the control scheme to adapt to inter-individual and intra-individual changes.

	\subsection{Linearization for Controller Design}
	
	A linear model with 12 states is derived by linearizing and reducing the nonlinear G\"ottingen Minipig model around a steady state BG concentration of 110 mg/dL, which is used in the controller design. The parameter $k_{\textrm{IS}}(t)$ enters the nonlinear system linearly and effects only one state. We therefore obtain the following time-varying plant
	\begin{equation}\label{eq:Linear_Model_time_varying}
	\begin{split}
	&\bm{\dot{x}}(t)=\bm{A}(k_{\textrm{IS}}(t))\bm{x}(t)+\bm{B}u(t)\\
	&y(t)=\bm{C}\bm{x}(t)
	\end{split}
	\end{equation}
	where $\bm{x}(t) \in \mathbb{R}^{12\times 1}$, $u(t) \in \mathbb{R}$, $y(t) \in \mathbb{R}$, $\bm{A}(k_{\textrm{IS}}(t))~\in~\mathbb{R}^{12\times 12}$, $\bm{B} \in \mathbb{R}^{12\times 1}$, $\bm{C} \in \mathbb{R}^{1\times 12}$ and only one entry in $\bm{A}(k_{\textrm{IS}}(t))$ is time-dependent.
	The input $u(t)$~[U] is the subcutaneous insulin injection per sampling period. The output is the blood glucose concentration [mg/dL].
	The training data for the GP is derived on basis of the linear time-varying model.
	For the MPC design, we fix the IS to its nominal value of 1~mg/(min$\cdot$mU) to obtain a linear time-invariant plant model.
	Due to space limitation, a detailed description of the linear model and matrices can be found in the appendix of \cite{Ortmann2017}.
	
	\section{Controller Design}\label{Sec:Controller}
	
\begin{figure}
		\centering
		{\includegraphics*[scale=.53,trim={5.2cm 10cm 12.8cm 3cm},clip]{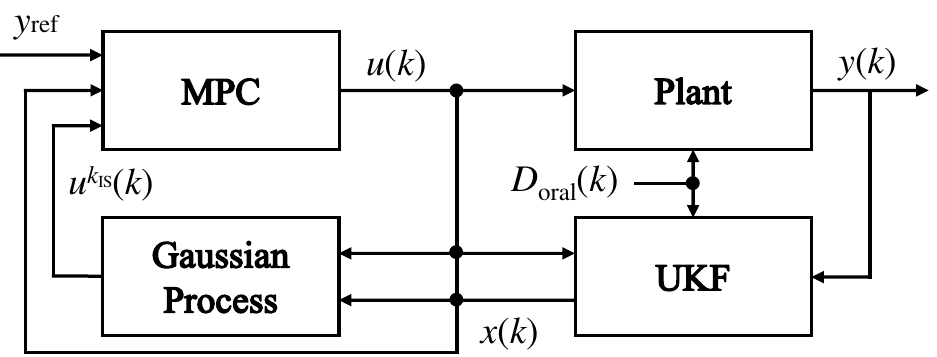}}
		\caption{Block diagram of the GP-MPC.}
		\label{fig:blockdiagram}
	\end{figure}

	The proposed controller utilizes an UKF, a GP and an MPC to counteract the problems arising from a changing IS. The UKF observes the states, which are used in the MPC and in calculating the training data for the GP. The GP predicts the future influence of the changing IS on the BG concentration and provides it to the MPC, which calculates the subcutaneous insulin input $u(k)$. The schematic of the proposed control scheme is provided in Fig. \ref{fig:blockdiagram}.
	\subsection{Unscented Kalman Filter}\label{SubSec:UKF}
	To estimate the states of the metabolic system, the UKF from \cite{Misgeld2017} is used. The inputs to the filter are the subcutaneous insulin input $u(k)$, the food intake $D_\textrm{oral}(k)$ and the intravenous BG measurement $y(k)$. The model used in the UKF is a reduced version of the G\"ottingen Minipig model that was derived in \cite{Misgeld2016a} using model residualization to neglect unobservable states. To calculate the means and covariances for the state update in the filter, sigma points are mapped through the reduced-order nonlinear model. The posterior mean and covariance are then approximated by a weighted sample mean and covariance of the posterior sigma points, respectively.
	The effect of food intake, through the gastro-intestinal tract, on the metabolic system is incorporated in the UKF through a feed-forward block. In particular, the gastro-intestinal tract is modelled by coupled linear first-order ordinary differential equations \cite{Hovorka2004}:
	\begin{equation}
	\begin{split}
	T_\textrm{{ing,1}}\cdot\dot{M}^\textrm{{G}}_\textrm{{ing,1}}(t)&=D_\textrm{{oral}}(t)-M^\textrm{{G}}_\textrm{{ing,1}}(t)\\
	T_\textrm{{ing,2}}\cdot\dot{M}^\textrm{{G}}_\textrm{{ing,2}}(t)&=M^\textrm{{G}}_\textrm{{ing,1}}(t)-M^\textrm{{G}}_\textrm{{ing,2}}(t)\\
	D_\textrm{{oral}}(t)&=d_\textrm{{oral}}\cdot\delta(t)\\
	r_\textrm{{GGA}}(t)&=f_\textrm{{G}}\cdot M^\textrm{{G}}_\textrm{{ing,2}}(t)
	\end{split}
	\end{equation}
	with time constants $T_\textrm{{ing,1}}$ and $T_\textrm{{ing,2}}$, the bioavailability of glucose $f_\textrm{{G}}$, the glucose amount $d_\textrm{{oral}}$ and the states $M^\textrm{{G}}_\textrm{{ing,1}}(t)$ and $M^\textrm{{G}}_\textrm{{ing,2}}(t)$ modelling the solid phase of the glucose mass flow in the stomach and the liquid phase in the intestine, respectively.
	The input to this subsystem is the announced exogenous glucose input $D_\textrm{{oral}}(t)$ and the output of this system is the glucose appearance
	rate in the blood stream $r_\textrm{{GGA}}(t)$.
	By including this gastro-intestinal tract model in the UKF, the state estimate provides the MPC with the information of an exogenous glucose disturbance and it can therefore counteract the food intake by increasing the amount of insulin infusion. Including the exogenous glucose input also enables one to attenuate meal-induced disturbances when learning the individual's circadian IS rhythm.
	
	\subsection{Gaussian Process}\label{SubSec:GP}
	A GP is a machine learning technique that can be used to predict future values of a process based on training data derived from historical measurements. It provides the user with an expected value and higher moments of the prediction. The core element of a GP is its kernel function. There exist a variety of standard kernel functions in the literature, each with its own purpose. The space of kernel functions is closed under multiplication so we can obtain customized kernel functions by multiplying them \cite{Rasmussen2006}. In \cite{Klenske2016}, a GP was used to predict future disturbances of a telescope mount to include these predictions in an MPC to improve the control performance. Following a similar approach, the GP in this paper is used to predict the future influence of the time-varying IS on the BG concentration.
	
	\subsubsection{Periodic Kernel Function}\label{SubSubSec:Periodic_Kernel_Function}
	
	The kernel function used in this paper combines a squared exponential kernel function $k_\textrm{{SE}}(t,t';l_\textrm{{SE}})$ and a periodic kernel function $k_\textrm{{P}}(t,t';l_\textrm{P},\lambda)$:  
	\begin{equation}
	k_\textrm{{SE}}(t,t';l_\textrm{{SE}})=\exp\left(-(t-t')^2/2l_{\textrm{SE}}^2\right)
	\end{equation}
	\begin{equation}
	k_\textrm{{P}}(t,t';l_\textrm{{P}},\lambda)=\exp\left(-2\sin^2\left(\dfrac{\pi}{\lambda}(t-t')\right)/l_\textrm{{P}}^2\right).
	\end{equation}
	The periodic part enables us to predict the influence of the periodic IS into the future, while the squared exponential part provides the possibility to relax the assumption of perfect periodicity.
	The resulting kernel function is given by
	\begin{equation}
	k_\textrm{{C}}(t,t';\bm{\eta})=\theta^2\cdot k_\textrm{{SE}}(t,t';l_\textrm{{SE}}) \cdot k_\textrm{{P}}(t,t';l_\textrm{P},\lambda)
	\end{equation}
	and has 4 hyperparameters, which are lumped in a vector $\bm{\eta}:=[\theta^2,l_\textrm{{SE}},l_\textrm{{P}},\lambda]$.
	Amongst these, $\theta^2$ is the variance, $l_\textrm{{SE}}$ enables us to tune the decaying trust in past training data, $l_\textrm{{P}}$ is a length-scale and $\lambda$ is the periodic-length of the training data. The hyperparameters $\theta$ and $l_\textrm{P}$ are tuned on the training data by maximizing the marginal likelihood
	\begin{equation}
	\log p(\bm{y},\bm{\eta})=-\dfrac{1}{2}(\bm{y}^T\bm{K}(\bm{\eta})^{-1}\bm{y}+\log\bm{K}(\bm{\eta})+n\log2\pi),
	\end{equation}
	where $\bm{y}$ are the training data points, $\bm{K}(\bm{\eta})$ is the kernel function $k_\textrm{C}(\bm{\eta})$ evaluated at the time stamps of $\bm{y}$ and $n$ is the number of training points. The initial values for the optimization are important, because fitting the hyperparameters is a non-convex optimization problem. They are chosen as $\sigma~=~1$~mg/dL and $l_\textrm{P}~=~1$, while ${\lambda=~24\cdot 60}$~min$~=~1440$~min and $l_\textrm{SE}~=~10^9$~min are fixed. The parameter $l_\textrm{SE}$ is chosen large to have no decay of trust in the training data as they are derived from the last 2.5 days.
	
	\subsubsection{Generating the Training Data}\label{SubSubSec:Generating_the_Training_Data}
		
	The training data for the GP are the disturbances $u^{k_{\textrm{IS}}}_{k-1}$, which are obtained during closed-loop control. With every new BG measurement a new disturbance value is calculated and added to the training data set.
	The training data reflects the difference of the changing IS from its nominal value and its effect on the BG concentration. We model this effect as a disturbance acting on a linear time-invariant plant with nominal IS. The training data consist of this disturbance signal and the corresponding time stamps.
	To model the influence of the changing IS as a disturbance, we separate the system matrix $\bm{A}(k_{\textrm{IS}}(t))$ into a time-invariant part $\bm{\hat{A}}$ and a time-varying part $\bm{A}^{k_{\textrm{IS}}}(t)$:
	\begin{equation}
	\begin{split}
	\bm{\dot{x}}(t)	&=\bm{A}(k_{\textrm{IS}}(t))\bm{x}(t)+\bm{B}u(t)\\
	&=\bm{\hat{A}}\bm{x}(t)+\bm{A}^{k_{\textrm{IS}}}(t)\bm{x}(t)+\bm{B}u(t).
	\end{split}
	\end{equation}
	Here, $\bm{\hat{A}}$ models the dynamics for the nominal IS of $k_{\textrm{IS,nom}}~=~1$~mg/(min$\cdot$mU) and $\bm{A}^{k_{\textrm{IS}}}(t)$ incorporates the dynamics due to the difference from the nominal IS.
	We then model $\bm{A}^{k_{\textrm{IS}}}(t)\bm{x}(t)$ as an IS-inducted disturbance $u^{k_{\textrm{IS}}}(t)$ entering the system through $\bm{B}^{k_{\textrm{IS}}}$ and get:
	\begin{equation}\label{eq:linear_system}
	\bm{\dot{x}}(t)=\bm{\hat{A}}\bm{x}(t)+\bm{B}u(t)+\bm{B}^{k_{\textrm{IS}}}u^{k_{\textrm{IS}}}(t).
	\end{equation}
	The vector $\bm{B}^{k_{\textrm{IS}}}$ has only one non-zero entry, because $\bm{A}^{k_{\textrm{IS}}}(t)$ has only one non-zero entry
	(see appendix).
	This entry links the insulin concentration in the interstitial of the muscular/adipose tissue to the change in BG concentration in the intracellular fluid space of muscular/adipose tissue (see (\ref{eq:MiniPig_k_IS})).
	To generate the training data, we discretise the system in (\ref{eq:linear_system}) with a sampling time of 5 minutes and obtain:
	\begin{equation}\label{Eq:Linear_Model}
	\bm{x}_{k+1}=\bm{\hat{A}}_{d}\bm{x}_{k}+\bm{B}_{d}u_{k}+\bm{B}^{k_{\textrm{IS}}}_{d}u^{k_{\textrm{IS}}}_{k}.
	\end{equation}
	We introduce the notation $[\cdot]_i$ to refer to the $i$th row of a vector/matrix variable and denote $[\bm{x}]_{i^*}$ as the row in $\bm{x}$ corresponding to the BG concentration in the vascular blood space of muscular/adipose tissue.
	At every time step, we calculate the IS-inducted disturbance using:
	\begin{equation}
	u^{k_{\textrm{IS}}}_{k-1}=\left(\left[\bm{x}_{k}\right]_{i^*}-\left[\bm{\hat{A}}_{d}\bm{x}_{k-1}\right]_{i^*}-\left[\bm{B}_{d}u_{k-1}\right]_{i^*}\right)/\left[\bm{B}^{k_{\textrm{IS}}}_{d}\right]_{i^*}.
	\end{equation}
	These calculated disturbances form the training data set of the GP.
	Through utilization of the UKF, which includes a model of the gastro-intestinal tract, food intake does not perturb our training data, if the food intake is correctly announced. The time of the food intake can therefore be arbitrary and the controller does not expect the same meal on the next day and will not give insulin even though it might not be needed.
	Once the training data are computed, they are collected and used to train the GP. At each time instant, the GP is trained using training data of 2.5 days, which has a filtering effect. Before performing the hyperparameter optimization on the training data, we apply a zero-phase filter. This filter erases noise and outliers in the training data, but prevents a phase shift so the predicted disturbance stays in phase with the metabolic process. The output of the GP is the disturbance $u^{k_{\textrm{IS}}}_{k}$, which is used in the MPC in \eqref{eq:MPC}.
	
	\subsection{Model Predictive Controller}\label{SubSec:MPC}
	
	We use an MPC to calculate the insulin input $u$ in terms of the amount of insulin units [U] delivered during each sample period (namely, 5 minutes). The information of the predicted disturbance from the GP is incorporated in the MPC's optimization problem. We use GP-MPC to denote the MPC using the information provided by the GP and will refer to the controller without the additional information as MPC.
	
	The GP-MPC and MPC both use the linear model in (\ref{Eq:Linear_Model}) that describes the plant around the linearization point and calculates the difference of the insulin injection from the basal value $u_\textrm{basal}$ that is used for the linearization. We limit our input to be positive, because we can not retrieve insulin from the body and the  counterregulatory hormone glucagon is not used in this control approach.
	The prediction of the future influence of the changing IS, which is provided by the GP, is incorporated into the system model as a disturbance $u^{k_{\textrm{IS}}}_k$. At time step $t$, we solve the following optimization problem using the Yalmip toolbox for Matlab \cite{Lofberg2004}:
	\begin{align}\label{eq:MPC}
	\begin{split}
	J^*_{0\rightarrow N-1}(&\bm{x}_0)=\min_{\bm{U}_{0\rightarrow N-1}} \;J_{0\rightarrow N-1}(\bm{x_0},\bm{U}_{0\rightarrow N-1})
	\\
	\text{s.t.} \; \bm{x}_{k+1}&=\bm{\hat{A}}_{d}\bm{x}_k+\bm{B}_{d}u_k+\bm{B}^{k_{\textrm{IS}}}_{d}u_k^{k_{\textrm{IS}}}, \;\;\bm{x}_0=\bm{x}(t)
	\\
	y_k&=\bm{C}_{d}\bm{x}_k,\;\; y_N = 0
	\\
	0 &\leq u_k+u_\textrm{basal}\leq u_\textrm{max}
	\end{split}
	\end{align}
	where the cost function is
	\begin{equation}
	J_{0\rightarrow N-1}(\cdot,\cdot)={\sum_{k=0}^{N-1}} y^T_{k}Qy_{k}+(u_{k}-u^{ss}_k)^TR(u_{k}-u^{ss}_k).
	\end{equation}
	Note that we introduce the vector $u^{ss}$ in the cost function to reject the disturbance $u^{k_{\textrm{IS}}}$. $u^{ss}_k$ is the input that is needed to reject the disturbance $u^{k_{\textrm{IS}}}_k$ at steady state and is calculated by solving the linear equations
	\begin{equation}
	\begin{bmatrix}
	\bm{\hat{A}}_d-\bm{I} & \bm{B}_d \\
	\bm{C}_d & 0
	\end{bmatrix}
	\begin{bmatrix}
	\bm{x}^{ss}_k \\
	u^{ss}_k
	\end{bmatrix} =	
	\begin{bmatrix}
	-\bm{B}^{k_{\textrm{IS}}}_du^{k_{\textrm{IS}}}_k \\
	0
	\end{bmatrix}
	\end{equation}
	$\forall k \in \left[0,N-1\right]$.
	After solving the optimization problem in (\ref{eq:MPC}), the first element of the input sequence $\bm{U}_{0\rightarrow N-1}$ is used and the optimization is repeated in a receding horizon manner every 5 minutes.
	
	The MPC is parametrized with ${N=30}$, ${Q=1}$ and ${R=40000}$. The prediction horizon is designed by considering the trade-off between calculation time of the optimization problem and the accuracy of the model prediction in the MPC. A longer control horizon would include more of the system dynamics, but would enlarge the number of decision variables in the optimization problem of the MPC. The control weight R is tuned for tracking performance and disturbance rejection. The design criteria is to have an undershoot after a food intake of 90g CHO of less than 10~mg/dL. At the same time the weight on the control input is supposed to lead to a controller that does not react too aggressively and does not overdose the patient with insulin, which could induce hypoglycemia. An upper bound $u_\textrm{max}$ on the insulin input of 0.5~U is introduced for the same reason. Also note, that to prevent insulin overdosing the linear 12 state model used in the MPC has 4 states, which keep track of the insulin in the body and its effect on the BG concentration.
	
	\section{Results}\label{Sec:Results}
	
	The nonlinear time-varying G\"ottingen Minipig model described in Section \ref{Sec:Modelling} is used to conduct \emph{in~silico} studies to validate the effectiveness of the proposed GP-MPC. We use the model parameters of Minpig 1 from \cite{Lunze2014a}, which has a basal insulin injection of $u_\textrm{basal} = 0.169$~U per sample period. 
	We first simulate fasting conditions in Subsection \ref{SubSec:fasting} to show that the GP-MPC has improved tracking performance over the MPC. We then consider the effect of announced meals in Subsection \ref{SubSec:announced}. For illustration purposes, we consider one meal a day and simulate announced meals of 50g CHO at 07:00h everyday to show that the GP-MPC is insensitive to food intake. Finally, we analyse the effects of a skipped meal in Subsection \ref{SubSec:skipped} to show that the GP-MPC is insensitive to changes in the amount or time of food intake. In each case, the simulation is performed for 7 days and the GP is activated after 2.5 days when enough training data is collected during closed-loop control, which is indicated by a vertical red dashed line in the figures. The statistics, for GP-MPC and MPC, are calculated using the data from that point onwards in time.
	
	The obtained results are provided in Figs. \ref{fig:fasting}-\ref{fig:skipped}. In these figures, panels A and C show the closed-loop behaviour for the MPC and GP-MPC for a period of 7 days. In panels B and D we show a close-up of Day 5 to analyse the control behaviour over one day. Panels A and B show the BG concentration, while panels C and D show the insulin injection. The yellow and green areas in panels A and B are the ``Safe Zone'' and ``Tight Zone'', which are defined as $\left[70, 180\right]$ mg/dL and $\left[80, 140\right]$ mg/dL, respectively. The reference for the controllers is a BG concentration of 110~mg/dL and the goal is to keep the BG in the ``Tight Zone'' to prevent hypoglycemia and hyperglycemia. The statistic results of the simulations can be found in Table \ref{tb:statistics}.
	
	\subsection{Fasting}\label{SubSec:fasting}
	
\begin{figure}
		\centering
		\hspace{-2cm}
		\subfigure{{\includegraphics*[scale=.308,trim={2.48cm 6.67cm 4.3cm 5.7cm},clip]{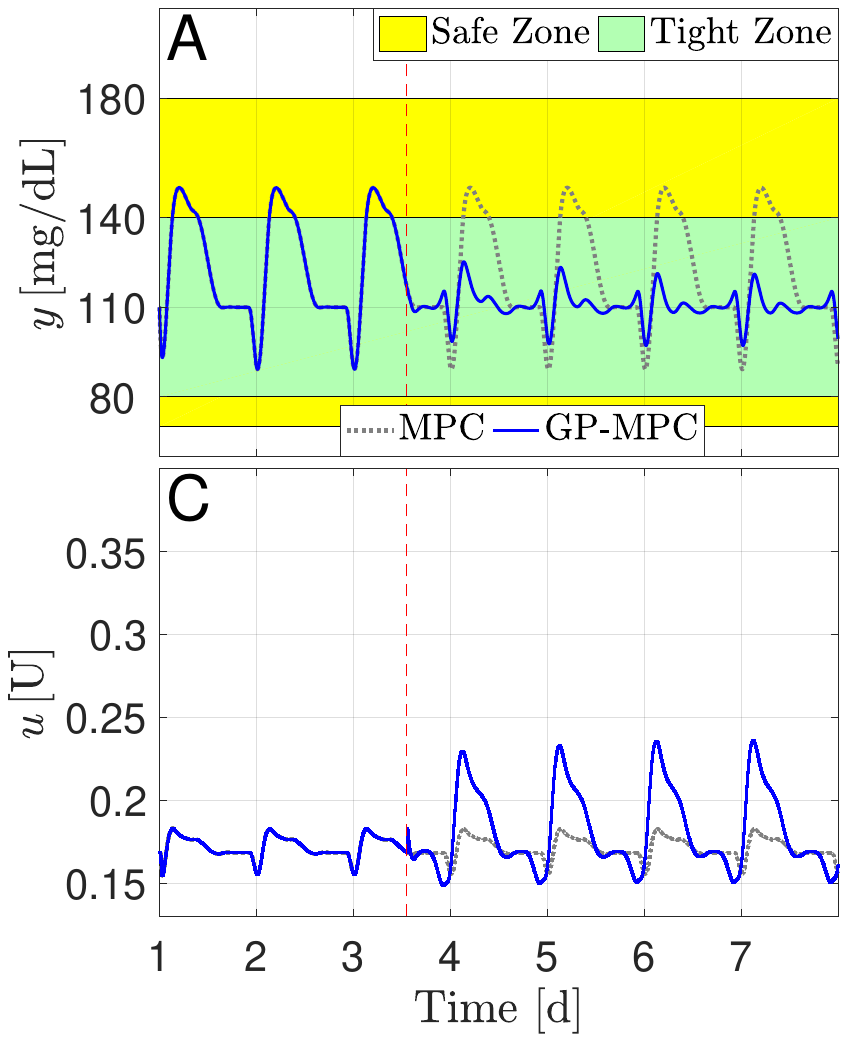}}
			\label{fig_first_case}}
		\hspace{-.41cm}
		\subfigure{{\includegraphics*[scale=.308,trim={4.75cm 6.67cm 2.3cm 5.7cm},clip]{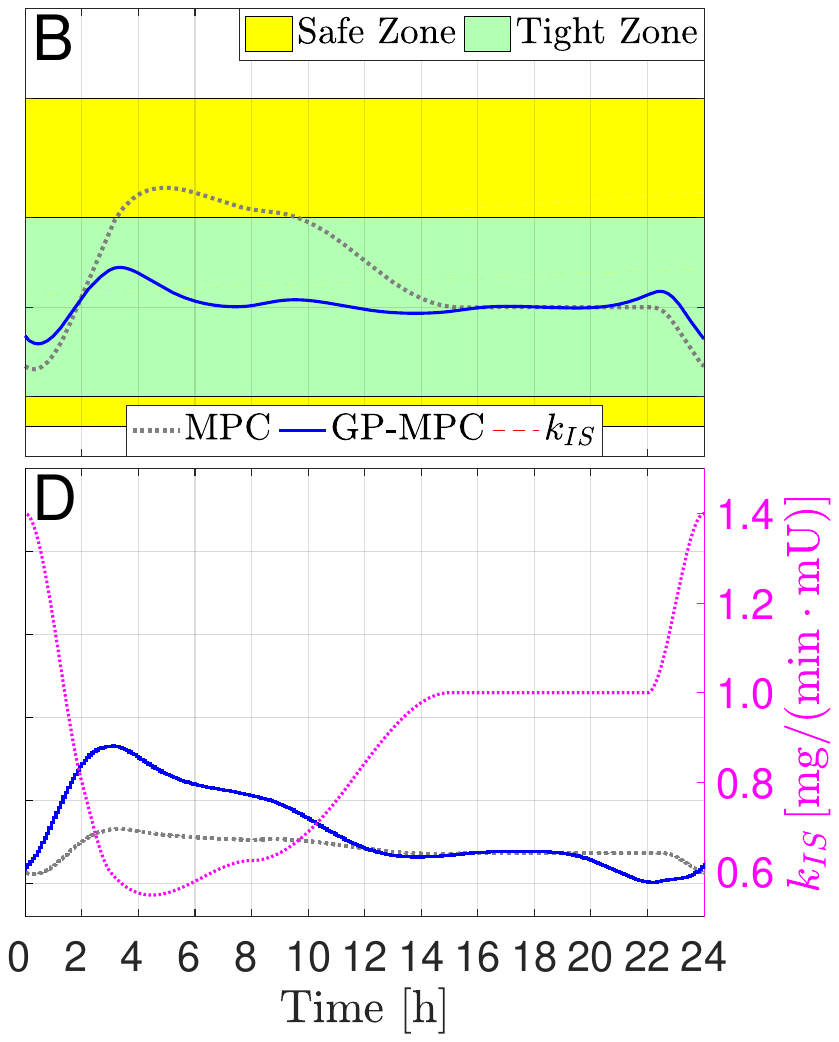}}
			\label{fig_second_case}}
		\hspace{-2.1cm}
		\caption{Fasting during closed-loop control with GP-MPC and MPC; circadian IS rhythm in panel D.}
		\label{fig:fasting}
	\end{figure}

	The fasting scenario in Fig. \ref{fig:fasting} shows the improved tracking performance of the GP-MPC in comparison to the MPC. Once enough training data is collected and the GP is activated, a reduction in tracking error can be observed. The mean BG concentration drops to 110.8~mg/dL in comparison to 120.5~mg/dL with the MPC. The standard deviation is also reduced, from 17.3~mg/dL to 4.6~mg/dL, and the GP-MPC is able to keep the BG in the ``Tight Zone'' at all times, whereas the MPC is controlling the BG concentration to the ``Tight Zone'' only 75.9\% of the time. The BG concentration at 07:00h in the morning is 110.4~mg/dL and closer to the set point than during control with the MPC where it is outside the ``Tight Zone'' at 145.3~mg/dL. Therefore the GP-MPC outperforms the MPC during breakfast time. In Fig. \ref{fig:fasting}D one can see how the controller is counteracting the changing IS. The insulin input of the GP-MPC (blue line) reacts inverse to the IS to compensate for the changes in the effectiveness of the insulin. Also note, that the insulin input of the GP-MPC decreases or increases before the IS changes, because the GP provides the MPC with a preview of the upcoming changes in the IS. For example, the insulin input is decreased by the GP-MPC around 19:00h to reduce the insulin in the body, because its effectiveness is going to increase from 22:00h onwards. This means the proposed control algorithm acts on the periodic behaviour predictively and does not only react to a difference between reference and measured BG concentration, like the MPC does.
	
	\subsection{Announced Meals}\label{SubSec:announced}
	
\begin{figure}
		\centering
		\hspace{-2.5cm}
		\subfigure{{\includegraphics*[scale=.327,trim={2.48cm 6.67cm 4.3cm 5.7cm},clip]{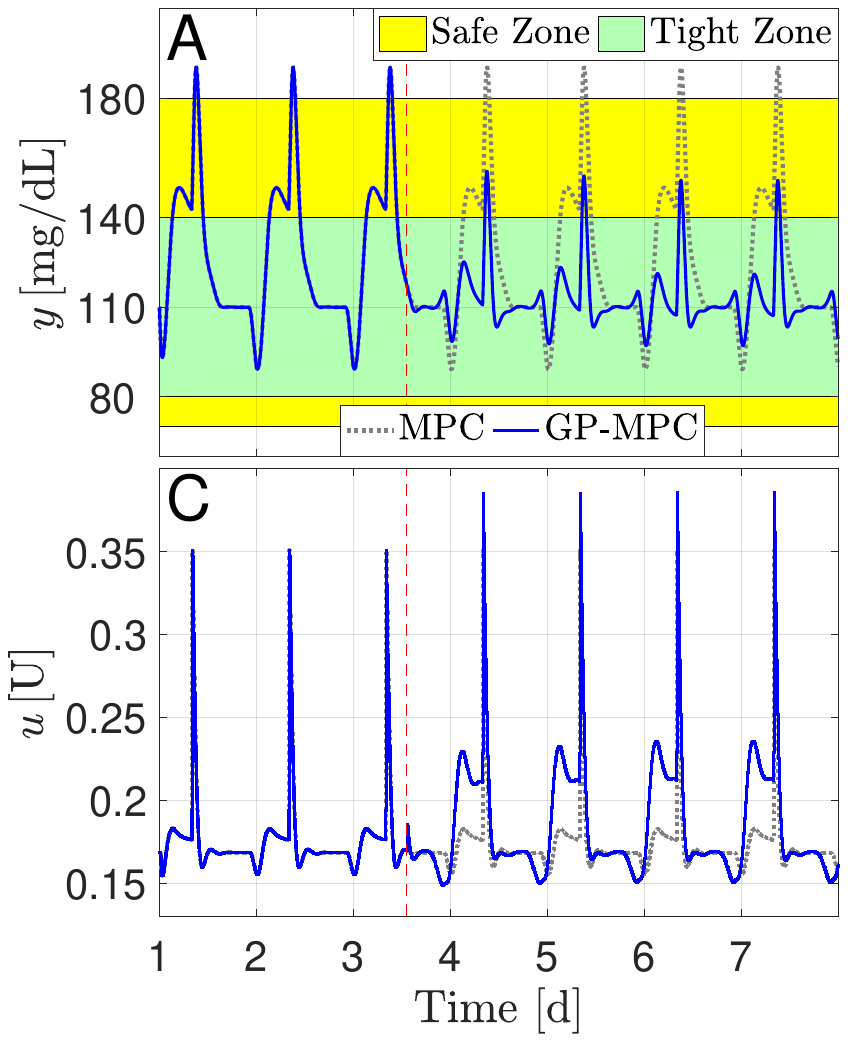}}
			\label{fig_first_case}}
		\hspace{-.41cm}
		\subfigure{{\includegraphics*[scale=.327,trim={4.75cm 6.67cm 4.1cm 5.7cm},clip]{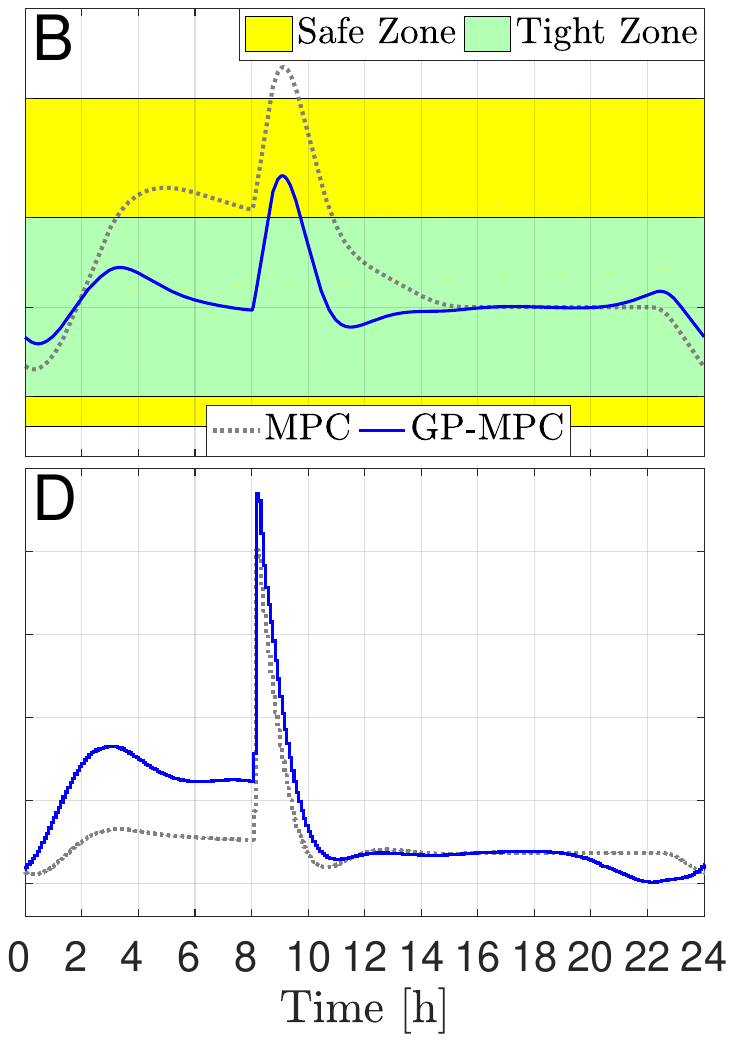}}
			\label{fig_second_case}}
		\hspace{-2.6cm}
		\caption{Announced meals during closed-loop control with GP-MPC and MPC.}
		\label{fig:announced}
	\end{figure}

	To show that the proposed control algorithm can reject disturbances (like meals) and that the training data calculation together with the GPs predictions are not disturbed by food intake, we analyse the control behaviour in the presence of announced meals. The amount of CHO and the time of food intake is announced to the UKF so the food intake can be included in the state estimation. Fig. \ref{fig:announced} shows that the food intake leads to a rapid rise in the BG concentration which needs to be reduced by increasing the insulin injection. For the MPC controller, the food intake raises the BG concentration out of the ``Safe Zone'' and therefore causes hyperglycemia during 3.8\% of the time. This is due to the high BG concentration at the time of food intake. Under control of the GP-MPC the BG concentration also has a peak, but due to the lower BG concentration at the time of food intake, there is no hyperglycemic event. The GP-MPC brings the BG concentration to the ``Tight Zone'' within less than two hours and the BG concentration around lunch time is also more desirable as compared to that of the MPC, because a lunch would again introduce a large peak in the BG concentration. Overall the GP-MPC profits from the better tracking performance with mean and standard deviation at 112.7~mg/dL and 9.9~mg/dL (against 123.8~mg/dL and 23.0~mg/dL for the MPC) and has improved \%-time in the ``Tight Zone'' with 95.6\% for the GP-MPC vs 71.3\% for the MPC. The results for the BG concentration at 07:00h are comparable to the fasting condition as the last meal has been digested by that time.
	\begin{table*}
		\centering
		\caption{Statistics of the \emph{in~silico} studies.}
		\label{tb:statistics}
		\begin{tabular}{llcccccc}
			\hline
			\hline
			& \multicolumn{1}{l}{} 					& mean BG $\pm$ SD 	& \%-time  	& \%-time   	& \%-time& \%-time   	& BG at 07:00h  \\
			& \multicolumn{1}{l}{} 					& (mg/dL)			&  	$<70$ mg/dL	&  	$\left[70-180\right]$ mg/dL		& $\left[80-140\right]$ mg/dL 					&   		$>180$ mg/dL			&  (mg/dL) \\
			\hline
			\multirow{2}{*}{Fasting}  & MPC             & 120.5 $\pm$ 17.3   	& 0                 & 100              	& 75.9          & 0                 & 145.3         \\ 
			& GP-MPC           	& 110.8 $\pm$ 4.6    	& 0                 & 100              	& 100 	        & 0                 & 110.4         \\
			\hline
			\multirow{2}{*}{Announced}& MPC             & 123.8 $\pm$ 23.0   	& 0                 & 96.2            	& 71.3          & 3.8               & 145.3         \\ 
			& GP-MPC           	& 112.7 $\pm$ 9.9    	& 0                 & 100              	& 95.6          & 0                 & 110.3         \\
			\hline
			\multirow{2}{*}{Skipped}  & MPC             & 123.0 $\pm$ 21.8   	& 0                 & 97.2             	& 72.5          & 2.9               & 145.3         \\
			& GP-MPC           	& 112.2 $\pm$ 9.2   	& 0                 & 100              	& 96.5          & 0                 & 110.3         \\
			\hline
			\hline
		\end{tabular}
	\end{table*}
	\subsection{Skipped Meals}\label{SubSec:skipped}
	
\begin{figure}
		\centering
		\hspace{-2.5cm}
		\subfigure{{\includegraphics*[scale=.327,trim={2.48cm 6.67cm 4.3cm 5.7cm},clip]{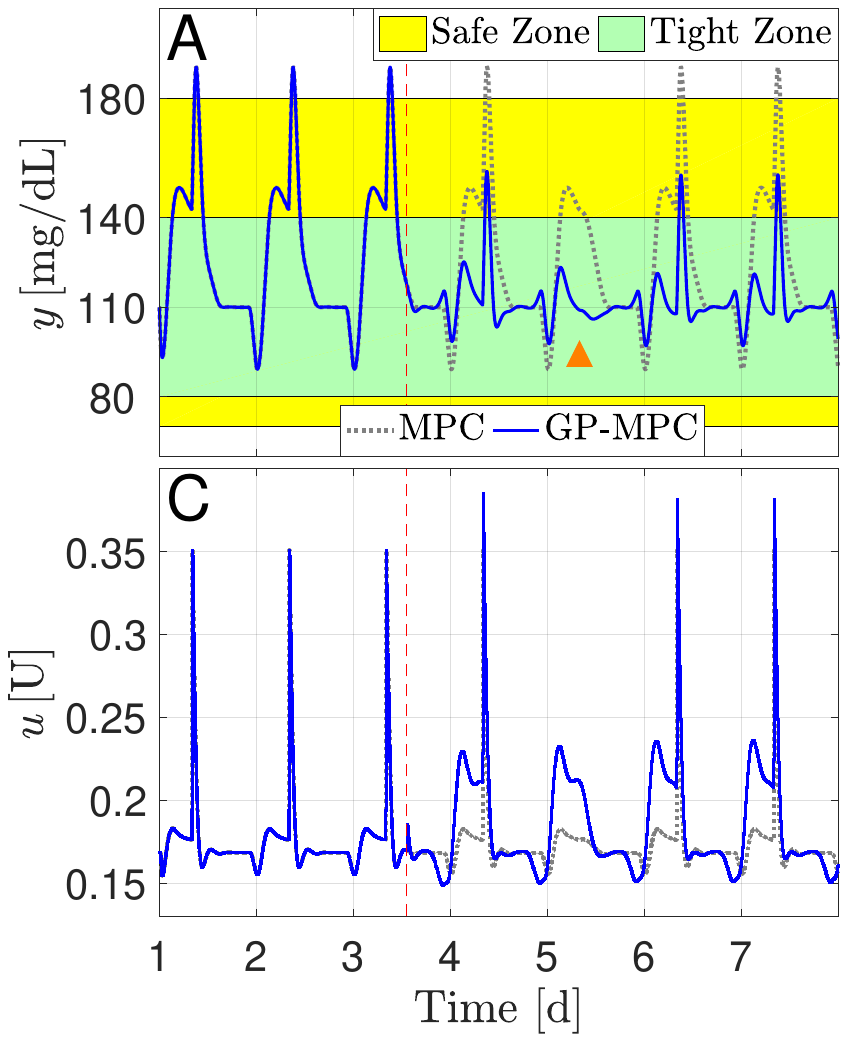}}
			\label{fig_first_case}}
		\hspace{-.41cm}
		\subfigure{{\includegraphics*[scale=.327,trim={4.75cm 6.67cm 4.1cm 5.7cm},clip]{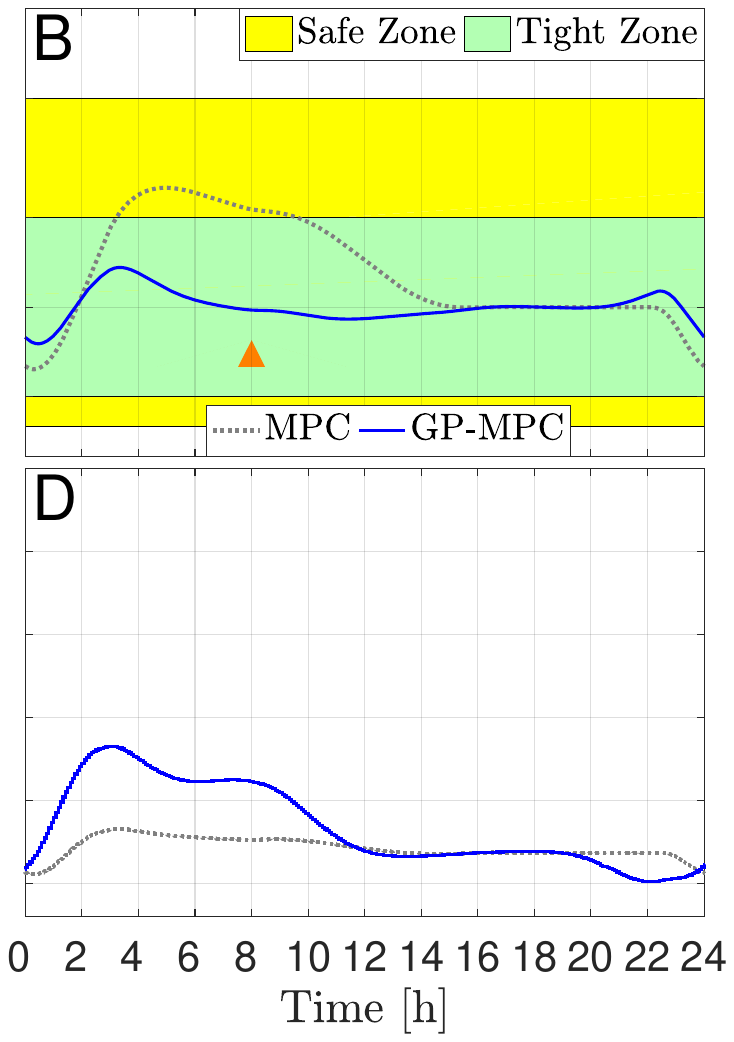}}
			\label{fig_second_case}}
		\hspace{-2.6cm}
		\caption{Announced meals with a skipped meal (marked with triangle) during closed-loop control with GP-MPC and MPC.}
		\label{fig:skipped}
	\end{figure}

	We will now present the GP-MPC's reaction to a skipped meal to show that the training data calculation is not disturbed by food intake. To do this, we simulate a skipped meal on the 5th day, which is marked with the orange triangle in Fig. \ref{fig:skipped}. The BG concentration around that time is not lowered by the GP-MPC, which means that the GP is not predicting a food intake. Slight changes in BG concentration occur due to the changing IS (see Fig. \ref{fig:fasting}D). This confirms that the amount and time of food intake can be arbitrary and do not alter the improvements in tracking performance of the GP-MPC in comparison to the MPC. This is also comfirmed by observing the mean BG concentration of 122.2~mg/dL, standard derivation of 9.2~mg/dL and \%-time in the tight range of 96.5\%, which differ only slightly from the results obtained in Subsection \ref{SubSec:announced}, caused by the missing peak of the skipped meal on the 5th day.

	\section{Conclusion}\label{Sec:Conclusion}
	
	In this work, a GP-MPC is proposed to counteract the problems arising from a changing IS. By utilizing a GP together with an UKF and an MPC, the proposed control algorithm is able to learn the effect of the individual's circadian IS rhythm during closed-loop control and is able to adapt to intra-individual changes of the metabolic system. The GP-MPC has a better tracking performance than the MPC and is insensitive to food intake, which therefore allows flexibility in choosing meal times and amounts.
	
	\bibliographystyle{IEEEtran}
	\bibliography{IEEEabrv,ASCC2017}
	
	\onecolumn
\begin{multicols}{2}
\appendix
Here, we provide more details on the linear time-varying model in (\ref{eq:Linear_Model_time_varying}). The input $u(t)$~[U] to the system is the subcutaneous insulin injection per sampling period. The output is the plasma glucose concentration [mg/dL]. The states of the 12 state linear time-variant model are: plasma glucose concentration [mg/dL], liver glucose concentration [mg/dL], glucose concentration in the vascular blood space of the muscle/adipose tissue [mg/dL], glucose concentration in the intracellular fluid space of the muscle/adipose tissue [mg/dL], insulin concentration in the interstitial space of muscle/adipose tissue [mU/L], normalized glucagon concentration in the liver [1], glucagon reduction/dissociation rate [1], normalized insulin concentration in the liver [1], non-monomeric (inactive) subcutaneous insulin [mU/min], monomeric (active) subcutaneous insulin [mU/min], glucose mass flow in the stomach [mg/min] and liquid phase in the intestine [mg/min]. The system matrices in (\ref{eq:Linear_Model_time_varying}) are:
\end{multicols}
\begin{multicols}{1}

\arraycolsep=1.2pt

\begin{equation*}
A(k_\textrm{IS}(t))=\left[\begin{array}{cccccccccccc} -1.14 & 0.494 & 0.647 & 0 & 0 & 0 & 0 & 0 & 0 & 0 & 0.0178 & 0\\ 3.68 & -4.56 & 0 & 0 & -0.018 & 86.5 & -96.8 & 59.3 & 0 & -0.073 & 0 & 0\\ 2.01 & 0 & -3.3 & 1.3 & 0 & 0 & 0 & 0 & 0 & 0 & 0 & 0\\ 0 & 0 & 0.2 & -0.2 & -0.1\cdot k_\textrm{IS}(t) & 0 & 0 & 0 & 0 & 0 & 0 & 0\\ 0 & 0 & 0 & 0 & -0.0697 & 0 & 0 & 0 & 0 & 0.0973 & 0 & 0\\ -0.0018 & 0 & 0 & 0 & -6.7\cdot 10^{-4} & -0.371 & 0 & 0 & 0 & -0.00272 & 0 & 0\\ 0 & 0 & 0 & 0 & 0 & 0.00687 & -0.0154 & 0 & 0 & 0 & 0 & 0\\ 0 & 0 & 0 & 0 & -2.08\cdot 10^{-4} & 0 & 0 & -0.04 & 0 & -8.42\cdot 10^{-4} & 0 & 0\\ 0 & 0 & 0 & 0 & 0 & 0 & 0 & 0 & -0.0166 & 0 & 0 & 0\\ 0 & 0 & 0 & 0 & 0 & 0 & 0 & 0 & 0.015 & -0.015 & 0 & 0\\ 0 & 0 & 0 & 0 & 0 & 0 & 0 & 0 & 0 & 0 & -0.027 & 0.027\\ 0 & 0 & 0 & 0 & 0 & 0 & 0 & 0 & 0 & 0 & 0 & -0.017 \end{array}\right],
\end{equation*}

\end{multicols}
\begin{multicols}{1}
\arraycolsep=4pt

\begin{align*}
\quad B = \left[\begin{array}{cccccccccccc} 0& 0& 0& 0& 0& 0& 0& 0& 2.23& 0.99& 0& 0 \end{array}\right]^T,
\quad
C = \left[\begin{array}{cccccccccccc} 1 & 0 & 0 & 0 & 0 & 0 & 0 & 0 & 0 & 0 & 0 & 0 \end{array}\right].
\end{align*}

\end{multicols}
\end{document}